\documentclass[preprint,superscriptaddress,aps]{revtex4}
\usepackage{amsfonts}
\usepackage{amssymb}
\usepackage{amsmath}
\usepackage{graphics}
\usepackage{graphicx}
\usepackage{color}
\usepackage[T1]{fontenc}
\usepackage{hyperref}
\newcommand{\bea}{\begin{eqnarray}}
\newcommand{\eea}{\end{eqnarray}}

\begin{document}
\title{Derivative expansion and the induced Chern-Simons term in $\mathcal{N}=1$, $d=3$ superspace}

\author{F. S. Gama}
\email{fgama@fisica.ufpb.br}
\affiliation{Departamento de F\'{\i}sica, Universidade Federal da Para\'{\i}ba\\
 Caixa Postal 5008, 58051-970, Jo\~ao Pessoa, Para\'{\i}ba, Brazil}

\author{J. R. Nascimento}
\email{jroberto@fisica.ufpb.br}
\affiliation{Departamento de F\'{\i}sica, Universidade Federal da Para\'{\i}ba\\
 Caixa Postal 5008, 58051-970, Jo\~ao Pessoa, Para\'{\i}ba, Brazil}

\author{A. Yu. Petrov}
\email{petrov@fisica.ufpb.br}
\affiliation{Departamento de F\'{\i}sica, Universidade Federal da Para\'{\i}ba\\
 Caixa Postal 5008, 58051-970, Jo\~ao Pessoa, Para\'{\i}ba, Brazil}

\begin{abstract}
In this paper we apply a supersymmetric generalization of the method of derivative expansion to compute the induced non-Abelian Chern-Simons term in $\mathcal{N}=1$, $d=3$ superspace, for an arbitrary gauge group.
\end{abstract}

\maketitle

\section{INTRODUCTION}

The three-dimensional field theory models represent themselves as a convenient laboratory for study of many sophisticated aspects of quantum field theory. The main reasons for this are, first, the simplicity of formulation of these theories, second, their one-loop finiteness. For the three-dimensional supersymmetric field theories, the most convenient description is the superfield one \cite{Superbooks} which essentially simplifies the perturbative calculations. The superfield approach allowed to obtain many interesting results for quantum corrections in these theories. The most important ones among them are the explicit calculation of the one- and two-loop effective potential in a general scalar superfield model \cite{ourEP} and in other superfield models \cite{ourEP1}, proof of the explicit all-loop finiteness of the three-dimensional superfield QED \cite{qedfin}, and explicit calculation of the gauge sector of one-loop effective action in theories with extended (${\cal N}=2$ and ${\cal N}=4$) supersymmetry \cite{BS}.

At the same time, an interesting problem related to these theories is the problem of an effective (emergent) dynamics. Following this idea, the known field theory models emerge as effective theories, whereas the role of the fundamental theory is played by some simple model involving coupling of a some matter (which is further integrated out thus being unobserved) with the physically interesting fields. For the gauge theories (including the supersymmetric ones), the paradigmatic example of the emergent dynamics is the arising of the Abelian Maxwell term in the $CP^{N-1}$ theory (see f.e. \cite{cpn}). The concept of the emergent gravity, discussed f.e. in \cite{Liberati}, plays an important role within this concept since it can allow for solving the problem of consistent description of quantum gravity. In this paper, we apply this methodology to generate the non-Abelian Chern-Simons term. While in the three-dimensional non-supersymmetric case it has been done in \cite{BDP,Dunne}, and in the four-dimensional Lorentz-breaking case -- in \cite{CSYM}, it has not been done in superfield theories up to now.

The structure of the paper looks like follows. In the section 2, we describe the derivative expansion methodology for supersymmetyric gauge theories. In the section 3, we explicitly generate the non-Abelian Chern-Simons term. The Summary is devoted to discussing results and perspectives.

\section{DERIVATIVE EXPANSION SCHEME}

Let us start with the following three-dimensional matter action
\bea
\label{classicalaction}
S[\Phi,\bar\Phi,A_\alpha]=\int d^5z\bar\Phi(\nabla^2+m)\Phi \ ,
\eea
where the massive complex scalar superfield $\Phi$ interacts with the background non-Abelian Lie-algebra valued gauge superfield $A_\alpha=A_{\alpha}^a T^a$ (with $T^a$ are Lie algebra generators) through the minimally-coupled gauge covariant derivative $\nabla_\alpha\equiv D_\alpha-iA_\alpha$. Here we are using the scaled gauge superfield $gA_\alpha\rightarrow A_\alpha$, where $g$ is the coupling constant.

From $S[\Phi,\bar\Phi,A_\alpha]$, we can compute the one-loop effective action $\Gamma_{eff}[A_\alpha]$ by
formally integrating out the complex scalar superfields. Therefore, we arrive at the following equation:
\bea
\label{quantumaction}
\Gamma_{eff}[A_\alpha]=-\textrm{Tr}\ln(\nabla^2+m)=-\textrm{Tr}\ln\big\{D^2-\frac{i}{2}\big[2A^\alpha D_\alpha+(D^\alpha A_\alpha)-iA^\alpha A_\alpha\big]+m\big\} \ .
\eea
Here, we use the usual definition of the trace operation for functional operators, namely
\bea
\label{functrace}
\textrm{Tr}\ln\mathcal{\hat O}=\textrm{tr}\int d^5z\ln\mathcal{O}\delta^5(z-z^\prime)|_{z=z^\prime} \ ,
\eea
where "tr" means trace over group indices, while $\delta^5(z-z^\prime)\equiv\delta^3(x-x^\prime)\delta^2(\theta-\theta^\prime)$.

It is convenient to split the effective action into two parts according to their parity under the transformation $m\rightarrow-m$,
\bea
\Gamma_{eff}[A_\alpha]&\equiv&\Gamma_{even}[A_\alpha]+\Gamma_{odd}[A_\alpha] \ ; \\
\label{even}
\Gamma_{even}[A_\alpha]&\equiv&-\frac{1}{2}\textrm{Tr}[\ln(\nabla^2+m)+\ln(\nabla^2-m)]=-\frac{1}{2}\textrm{Tr}\ln(\Box_A-iW^\alpha\nabla_\alpha-m^2) \ ,\\
\label{odd}
\Gamma_{odd}[A_\alpha]&\equiv&-\frac{1}{2}\textrm{Tr}[\ln(\nabla^2+m)-\ln(\nabla^2-m)] \ .
\eea
In a low-energy approximation or, equivalently, in an approximation of slowly varying background superfields, the even-parity part (\ref{even}) corresponds to the Euler-Heisenberg effective action and the odd-parity part (\ref{odd}) corresponds to the Chern-Simons action \cite{Lee}. In Ref. \cite{Red}, the one-loop effective action were calculated in three-dimensional Minkowski space. The supersymmetric generalization of these results was studied in \cite{BS}, in the Abelian case, to the $\mathcal{N}=2$ and $\mathcal{N}=4$ superspaces.

In \cite{BDP}, the authors applied the derivative-expansion scheme proposed in \cite{AF} to compute the induced  Chern-Simons term at zero and finite temperatures. The main goal of the present paper is to provide a systematic method to compute the non-Abelian Chern-Simons term induced by radiative corrections in $\mathcal{N}=1$, $d=3$ superspace via a derivative-expansion scheme so that, our results can be seen as a supersymmetric generalization of the result obtained in \cite{BDP} at zero temperature. In particular, we will show explicitly that, in low energies, the odd-parity part (\ref{odd}) corresponds to the Chern-Simons action, namely \cite{Superbooks}
\bea
\label{ChernSimons}
\Gamma_{odd}[A_\alpha]=C \ \textrm{tr}\int d^5z\big(A^\alpha W_\alpha+\frac{i}{6}\{A^\alpha,A^\beta\}D_\beta A_\alpha+\frac{1}{12}\{A^\alpha,A^\beta\}\{A_\alpha, A_\beta\}\big) \ ,
\eea
where $C$ is a dimensionless constant to be determined, and
\bea
\label{strengthfield}
W_\alpha=\frac{1}{2}D^\beta D_\alpha A_\beta-\frac{i}{2}[A^\beta,D_\beta A_\alpha]-\frac{1}{6}[A^\beta,\{A_\beta,A_\alpha\}] \ .
\eea

In order to apply the method of derivative expansion, let us substitute (\ref{quantumaction}) in (\ref{odd}), so that we can rewrite the result as
\bea
\label{odd2}
\Gamma_{odd}[A_\alpha]&=&-\frac{1}{2}\textrm{Tr}\bigg\{\ln\Big[1-\frac{i}{2}\big(2A^\alpha D_\alpha+(D^\alpha A_\alpha)-iA^\alpha A_\alpha\big)\frac{D^2-m}{\Box-m^2}\Big]\nonumber \\
&-&\ln\Big[1-\frac{i}{2}\big(2A^\alpha D_\alpha+(D^\alpha A_\alpha)-iA^\alpha A_\alpha\big)\frac{D^2+m}{\Box-m^2}\Big]\bigg\}+\Gamma_{odd}[A_\alpha=0] \ .
\eea
Since the inverse of the $A_\alpha$-propagator, namely $\Gamma_{odd}[A_\alpha=0]$, does not depend on the background superfield, it follows that we can drop it out by means of the normalization of the effective action.

Expanding the logarithmic terms of (\ref{odd2}), we get
\bea
\label{odd3}
\Gamma_{odd}[A_\alpha]=\sum_{n=1}^\infty\Big(\frac{i}{2}\Big)^n\frac{S^{(n)}}{n} \ ,
\eea
where
\bea
\label{taylor}
S^{(n)}&\equiv&\frac{1}{2}\textrm{Tr}\bigg\{\Big[\big(2A^\alpha D_\alpha+(D^\alpha A_\alpha)-iA^\alpha A_\alpha\big)\frac{D^2-m}{\Box-m^2}\Big]^n-\Big[\big(2A^\alpha D_\alpha+(D^\alpha A_\alpha)\nonumber \\
&-&iA^\alpha A_\alpha\big)\frac{D^2+m}{\Box-m^2}\Big]^n\bigg\} \ .
\eea
Notice that the derivatives $D_\alpha$ and $\partial_{\alpha\beta}$ acts on all functions at its right side, including the delta function (\ref{functrace}). For the purpose of calculating the functional trace in (\ref{taylor}), we need to move all derivatives $D_\alpha$ and $\partial_{\alpha\beta}$ to the right and all functions of $A_\alpha$ and their derivatives to the left so that, at the end of the calculation, we can calculate the functional trace by using the Fourier representation of the delta function. In order to perform this task, we need to use the following identities:
\bea
\label{Leibniz}
&&[D_\alpha,f(z)\}=(D_\alpha f)\Longleftrightarrow D_\alpha f(z)=(D_\alpha f)+(-)^{\epsilon_f}f D_\alpha  \ , \\
\label{commutators}
&&\frac{1}{\Box-m^2}f(z)=f\frac{1}{\Box-m^2}-[\Box,f]\frac{1}{(\Box-m^2)^2}+[\Box,[\Box,f]]\frac{1}{(\Box-m^2)^3}+\cdots ,
\eea
where $[ \ , \ \}$ is the graded commutator, while $f(z)$ is a function of background superfield and their derivatives. Notice that (\ref{Leibniz}) is the graded Leibniz rule.

Our calculation is also based on an extensive use of the following identities satisfied by the covariant derivatives $D_\alpha$:
\bea
\label{covariantder}
D_\alpha D_\beta=i\partial_{\alpha\beta}+C_{\beta\alpha}D^2 \ , \ \{D^2,D_\alpha\}=0 \ , \ [D_\alpha,D_\beta]=-2C_{\alpha\beta}D^2 \ , \ (D^2)^2=\Box \ .
\eea
Now, let us calculate the terms $S^{(n)}$ in (\ref{taylor}). However, taking into account that (\ref{ChernSimons}) is a functional which depends only on quadratic, cubic, and quartic terms in the background superfield, it follows that we only need to calculate the lowest order terms in the expansion of $\Gamma_{odd}$ in (\ref{odd3}), namely
\bea
\label{odd4}
\Gamma_{odd}[A_\alpha]=\frac{i}{2}S^{(1)}_{AA}-\frac{1}{8}(S^{(2)}_{AA}+S^{(2)}_{AAA}+S^{(2)}_{AAAA})-\frac{i}{24}(S^{(3)}_{AAA}+S^{(3)}_{AAAA})+\frac{1}{64}S^{(4)}_{AAAA} \ .
\eea
In the next section, we perform explicit calculations of all the terms in the expansion above.

\section{INDUCED CHERN-SIMONS TERM}

Let us start the calculation of the contributions $S^{(n)}$ in (\ref{odd4}). At first order, $n=1$, it is only necessary to calculate $S^{(1)}_{AA}$. Hence, it follows from (\ref{taylor}) that
\bea
\label{T1AA}
S^{(1)}_{AA}=im\textrm{Tr}\Big[A^\alpha A_\alpha\frac{1}{\Box-m^2}\Big]=im \ \textrm{tr}\int d^5z\Big[A^\alpha A_\alpha\frac{1}{\Box-m^2}\Big]\delta^5(z-z^\prime)|_{z=z^\prime} \ .
\eea
Since there is no covariant derivative $D_\alpha$ in the Eq. (\ref{T1AA}) and the Grassmann delta function satisfies the following identities:
\bea
\label{delta}
\delta^2(\theta-\theta^\prime)|_{\theta=\theta^\prime}=0 \ , \ D_\alpha\delta^2(\theta-\theta^\prime)|_{\theta=\theta^\prime}=0  \ , \ D^2\delta^2(\theta-\theta^\prime)|_{\theta=\theta^\prime}=1 \ .
\eea
It follows that (\ref{T1AA}) vanishes identically:
\bea
\label{1AA}
S^{(1)}_{AA}=0 \ .
\eea
Let us move on and calculate the second-order contributions in (\ref{odd4}), namely $S^{(2)}_{AA}$, $S^{(2)}_{AAA}$, and $S^{(2)}_{AAAA}$. First, let us begin with the quadratic contribution in the background superfield:
\bea
\label{T2AA}
S^{(2)}_{AA}&=&-m\textrm{Tr}\bigg[X_A\frac{D^2}{\Box-m^2}X_A\frac{1}{\Box-m^2}+X_A\frac{1}{\Box-m^2}X_A\frac{D^2}{\Box-m^2}\bigg] \ ;\\
X_A&\equiv&2A^\alpha D_\alpha+(D^\alpha A_\alpha) \ ,
\eea
where again we have used Eq. (\ref{taylor}). By assuming that the background superfield varies slowly in superspace, we can discard terms involving spinor derivatives higher than the second order acting on $A_\alpha$ in the calculation of (\ref{T2AA}). Moreover, in this approximation, we can move the operator $(\Box-m^2)^{-1}$ to the right using the identity (\ref{commutators}), so that only the first two terms on the right of Eq.(\ref{commutators}) need be retained. Therefore, it follows from (\ref{commutators}) that
\bea
\label{com2AA}
\frac{1}{\Box-m^2}X_A&\approx&X_A\frac{1}{\Box-m^2}-\big[4(D^2A^\alpha)D^2D_\alpha-2(D^\lambda D^\gamma A^\alpha)D_\lambda D_\gamma D_\alpha\big]\frac{1}{(\Box-m^2)^2} \ ,
\eea
where we have used the commutators
\bea
&&[\Box , X_A]=[(D^2)^2 , X_A]\approx4(D^2A^\alpha)D^2D_\alpha-2(D^\lambda D^\gamma A^\alpha)D_\lambda D_\gamma D_\alpha \ ;\\
&&\big[\Box,[\Box ,X_A]\big]=\big[(D^2)^2,[(D^2)^2 , X_A]\big]\approx0 \ .
\eea
Moreover, neglecting terms involving derivatives higher than the second, we notice that
\bea
\label{anti}
\{D^2 \ , \ 4(D^2A^\alpha)D^2D_\alpha-2(D^\lambda D^\gamma A^\alpha)D_\lambda D_\gamma D_\alpha\}\approx0 \ .
\eea
Substituting (\ref{com2AA}) into (\ref{T2AA})  and using Eq. (\ref{anti}), we obtain
\bea
\label{T2AA'}
S^{(2)}_{AA}=-m\textrm{Tr}\bigg[\big(X_AD^2X_A+X_AX_AD^2\big)\frac{1}{(\Box-m^2)^2}\bigg] \ .
\eea
As a next step, we need to push all derivatives $D_\alpha$ to the right by means of Eq. (\ref{Leibniz}). This task can be carried out by using the expressions:
\bea
\label{XDX}
X_AD^2X_A&\approx&-\big[4A^\alpha(D^2A_\alpha)+4A^\alpha(D_\alpha D^\beta A_\beta)+(D^\alpha A_\alpha)(D^\beta A_\beta)-4A^\alpha A^\beta i\partial_{\alpha\beta}\big]D^2 \ ;\\
\label{XXD}
X_AX_AD^2&\approx&[2A^\alpha(D_\alpha D^\beta A_\beta)+(D^\alpha A_\alpha)(D^\beta A_\beta)-4A^\alpha A^\beta i\partial_{\alpha\beta}]D^2 \ .
\eea
where we have kept only terms which give a non-vanishing contribution (see Eq. (\ref{delta})).

Substituting (\ref{XDX}) and (\ref{XXD}) into (\ref{T2AA'}), and noticing that
\bea
\label{integral}
\frac{D^2}{(\Box-m^2)^2}\delta^5(z-z^\prime)|_{z=z^\prime}=\frac{1}{(\Box-m^2)^2}\delta^3(x-x^\prime)|_{x=x^\prime}=\frac{1}{8\pi|m|} \ ,
\eea
we obtain
\bea
\label{2AA}
S^{(2)}_{AA}=\frac{2m}{8\pi|m|}\textrm{tr}\int d^5zA^\alpha(D^\beta D_\alpha A_\beta) \ ,
\eea
where we have used the Fourier representation of the delta function in (\ref{integral}) and the identities (\ref{covariantder}). It is worth noting that the expression (\ref{2AA}) is, in its functional structure, similar to the quadratic term in (\ref{ChernSimons}). In particular, if $A_\alpha$ is an Abelian superfield, then (\ref{2AA}) is invariant under the gauge transformation $\delta A_\alpha=D_\alpha K$. Hence, in this particular case, if we substituted (\ref{1AA}) and (\ref{2AA}) in (\ref{odd4}), we should obtain the induced Abelian Chern-Simons action.

Next, let us calculate the cubic contribution in the background superfield, namely $S^{(2)}_{AAA}$. Due to the low-energy approximation, the terms involving spinor derivatives higher than the first order acting on $A_\alpha$ can be neglected. It follows that, at this approximation,
\bea
\label{iden}
[\Box , X_A]=[(D^2)^2 , X_A]\approx0 \ , \ [\Box , A^\alpha A_\alpha]=[(D^2)^2 , A^\alpha A_\alpha]\approx0 \ .
\eea
Therefore, it follows from (\ref{taylor}) that
\bea
\label{T2AAA'}
S^{(2)}_{AAA}=im\textrm{Tr}\bigg[\big(X_AA^\beta A_\beta D^2+A^\alpha A_\alpha X_AD^2+X_AD^2A^\beta A_\beta+A^\alpha A_\alpha D^2X_A\big)\frac{1}{(\Box-m^2)^2}\bigg] ,
\eea
where we have pushed the operator $(\Box-m^2)^{-1}$ to the right by using (\ref{commutators}) and (\ref{iden}). Again, each term in (\ref{T2AAA'}) can be calculated by moving all derivatives $D_\alpha$ to the right by means of Eq. (\ref{Leibniz}). Hence, the terms that give a non-vanishing contribution are
\bea
\label{32}
&&X_AA^\beta A_\beta D^2\approx[2A^\alpha(D_\alpha A^\beta)A_\beta-2A^\alpha A^\beta(D_\alpha A_\beta)+(D^\alpha A_\alpha)A^\beta A_\beta]D^2 \ ;\\
&&A^\alpha A_\alpha X_AD^2\approx A^\alpha A_\alpha(D^\beta A_\beta)D^2 \ \ ; \ \ X_AD^2A^\beta A_\beta\approx(D^\alpha A_\alpha)A^\beta A_\beta D^2 \ ;\\
\label{34}
&&A^\alpha A_\alpha D^2X_A\approx-A^\alpha A_\alpha(D^\beta A_\beta)D^2 \ .
\eea
Substituting (\ref{32}-\ref{34}) into (\ref{T2AAA'}), and calculating the functional trace, we get
\bea
\label{T2AAA''}
S^{(2)}_{AAA}=\frac{2im}{8\pi|m|}\textrm{tr}\int d^5z\big[(D^\alpha A_\alpha)A^\beta A_\beta+A^\alpha(D_\alpha A^\beta)A_\beta-A^\alpha A^\beta(D_\alpha A_\beta)\big] \ .
\eea
Notice that (\ref{T2AAA''}) is a surface term, which for present purposes, we may neglect. Therefore,
\bea
\label{T2AAA}
S^{(2)}_{AAA}=0 \ .
\eea
Finally, regarding the second-order contributions, let us move on and calculate $S^{(2)}_{AAAA}$. In this case, we discard terms involving covariant derivatives $D_\alpha$ of any order acting on the background superfield. Hence, it follows from (\ref{taylor}) that
\bea
\label{T2AAAA'}
S^{(2)}_{AAAA}&=&m\textrm{Tr}\bigg[A^\alpha A_\alpha\frac{D^2}{\Box-m^2}A^\beta A_\beta\frac{1}{\Box-m^2}+A^\alpha A_\alpha\frac{1}{\Box-m^2}A^\beta A_\beta\frac{D^2}{\Box-m^2}\bigg]\nonumber\\
&=&2m\textrm{Tr}\bigg[A^\alpha A_\alpha A^\beta A_\beta\frac{D^2}{(\Box-m^2)^2}\bigg] \ .
\eea
Therefore, using Eqs. (\ref{functrace}) and (\ref{integral}), we obtain
\bea
\label{T2AAAA}
S^{(2)}_{AAAA}=\frac{2m}{8\pi|m|}\textrm{tr}\int d^5zA^\alpha A_\alpha A^\beta A_\beta \ .
\eea
Let us now consider the calculation of the third-order contributions in (\ref{odd4}): $S^{(3)}_{AAA}$ and $S^{(3)}_{AAAA}$. First, let us start with the cubic contribution in the background superfield $S^{(3)}_{AAA}$. In particular, by means of the same reasoning used to obtain (\ref{T2AAA'}), we can use Eq. (\ref{taylor}) and write $S^{(3)}_{AAA}$ as
\bea
\label{T3AAA'}
S^{(3)}_{AAA}&=&-m\textrm{Tr}\bigg[\big(X_AD^2X_AX_AD^2+X_AX_AD^2X_AD^2+X_AD^2X_AD^2X_A\nonumber\\
&+&m^2X_AX_AX_A\big)\frac{1}{(\Box-m^2)^3}\bigg] \ .
\eea
As previously described, we need to push all derivatives $D_\alpha$ to the right using the identity (\ref{Leibniz}) and keep only terms which give a non-vanishing contribution. After some tedious algebraic work, we find that
\bea
\label{40}
X_AD^2X_AX_AD^2&\approx&8\big[A^\alpha(D^\beta A^\gamma)A^\lambda-A^\alpha A^\gamma(D^\beta A^\lambda)\big](-\partial_{\alpha\beta}\partial_{\gamma\lambda}+C_{\beta\alpha}C_{\lambda\gamma}\Box)D^2\nonumber\\
&+&4\big[2\{A^\alpha, A^\beta\}(D_\beta A_\alpha)-2A^\alpha(D_\alpha A^\beta)A_\beta+A^\alpha A_\alpha(D^\beta A_\beta)\nonumber\\
&+&A^\alpha(D^\beta A_\beta)A_\alpha-(D^\alpha A_\alpha)A^\beta A_\beta\big]\Box D^2 \ ;\\
X_AX_AD^2X_AD^2&\approx&8A^\alpha A^\beta(D^\gamma A^\lambda)(-\partial_{\alpha\beta}\partial_{\gamma\lambda}+C_{\beta\alpha}C_{\lambda\gamma}\Box)D^2+4\big[2A^\alpha(D_\alpha A^\beta)A_\beta\nonumber\\
&+&2[A^\alpha, A^\beta](D_\beta A_\alpha)-A^\alpha A_\alpha(D^\beta A_\beta)+A^\alpha(D^\beta A_\beta)A_\alpha\nonumber\\
&+&(D^\alpha A_\alpha)A^\beta A_\beta\big]\Box D^2 \ ;\\
X_AD^2X_AD^2X_A&\approx&-8\big[A^\alpha(D^\beta A^\gamma)A^\lambda+A^\alpha A^\beta(D^\gamma A^\lambda)-A^\alpha A^\gamma(D^\beta A^\lambda)\big](-\partial_{\alpha\beta}\partial_{\gamma\lambda}\nonumber\\
&+&C_{\beta\alpha}C_{\lambda\gamma}\Box)D^2+4\big[2A^\alpha(D_\alpha A^\beta)A_\beta-2\{A^\alpha, A^\beta\}(D_\beta A_\alpha)\nonumber\\
&+&A^\alpha A_\alpha(D^\beta A_\beta)-A^\alpha(D^\beta A_\beta)A_\alpha+(D^\alpha A_\alpha)A^\beta A_\beta\big]\Box D^2 \ ;\\
\label{43}
X_AX_AX_A&\approx&-4\big[2A^\alpha(D_\alpha A^\beta)A_\beta+2[A^\alpha,A^\beta](D_\beta A_\alpha)+A^\alpha A_\alpha(D^\beta A_\beta)\nonumber\\
&+&A^\alpha(D^\beta A_\beta)A_\alpha+(D^\alpha A_\alpha)A^\beta A_\beta\big]D^2 \ .
\eea
Hence, inserting Eqs. (\ref{40}-\ref{43}) into (\ref{T3AAA'}) and carrying out some algebraic manipulation, we obtain a very simple result:
\bea
\label{T3AAA}
S^{(3)}_{AAA}=-\frac{4m}{8\pi|m|}\textrm{tr}\int d^5z\{A^\alpha,A^\beta\}(D_\beta A_\alpha) \ .
\eea
Next, let us calculate the quartic contribution in the background superfield, namely $S^{(3)}_{AAAA}$. In an approximation of slowly varying background superfields, we neglect terms involving derivatives acting on the background superfield, just as was done in (\ref{T2AAAA'}). Therefore, from Eq. (\ref{taylor}), we obtain
\bea
\label{T3AAAA'}
S^{(3)}_{AAAA}&=&4im\textrm{Tr}\bigg[\big(A^\alpha D_\alpha D^2 A^\beta D_\beta A^\gamma A_\gamma D^2+A^\alpha D_\alpha D^2A^\beta A_\beta A^\gamma D_\gamma D^2+A^\alpha A_\alpha D^2\nonumber\\
&\times&A^\beta D_\beta A^\gamma D_\gamma D^2+A^\alpha D_\alpha A^\beta D_\beta D^2A^\gamma A_\gamma D^2+A^\alpha D_\alpha A^\beta A_\beta D^2A^\gamma D_\gamma D^2 \nonumber\\
&+&A^\alpha A_\alpha A^\beta D_\beta D^2A^\gamma D_\gamma D^2+A^\alpha D_\alpha D^2A^\beta D_\beta D^2A^\gamma A_\gamma+A^\alpha D_\alpha D^2A^\beta A_\beta D^2 \nonumber\\
&\times&A^\gamma D_\gamma+A^\alpha A_\alpha D^2A^\beta D_\beta D^2A^\gamma D_\gamma+m^2A^\alpha D_\alpha A^\beta D_\beta A^\gamma A_\gamma+m^2A^\alpha D_\alpha A^\beta A_\beta\nonumber\\
&\times&A^\gamma D_\gamma+m^2A^\alpha A_\alpha A^\beta D_\beta A^\gamma D_\gamma \big)\frac{1}{(\Box-m^2)^3}\bigg]\nonumber\\
&=&12im\textrm{Tr}\bigg[A^\alpha A_\alpha A^\beta A_\beta\frac{D^2}{(\Box-m^2)^2}\bigg] \ .
\eea
Hence, calculating the trace of (\ref{T3AAAA'}), we get
\bea
\label{T3AAAA}
S^{(3)}_{AAAA}=\frac{12im}{8\pi|m|}\textrm{tr}\int d^5zA^\alpha A_\alpha A^\beta A_\beta \ .
\eea
The last contribution which is needed to be calculated is $S^{(4)}_{AAAA}$. However, we will not calculate explicitly $S^{(4)}_{AAAA}$, because the calculation goes in the same way as the results obtained (\ref{T2AAAA}) and (\ref{T3AAAA}). Therefore, it is given by
\bea
\label{T4AAAA}
S^{(4)}_{AAAA}=0 \ .
\eea
Finally, substituting Eqs. (\ref{1AA}), (\ref{2AA}), (\ref{T2AAA}), (\ref{T2AAAA}), (\ref{T3AAA}), (\ref{T3AAAA}) and (\ref{T4AAAA}) into Eq. (\ref{odd4}), we obtain the induced non-Abelian Chern-Simons term in its most simple form
\bea
\label{SF}
\Gamma_{odd}[A_\alpha]=-\frac{1}{8\pi|m|}\textrm{tr}\int d^5z\frac{m}{2}\Big[\frac{1}{2}A^\alpha(D^\beta D_\alpha A_\beta)-\frac{i}{3}\{A^\alpha,A^\beta\}(D_\beta A_\alpha)-\frac{1}{2}A^\alpha A_\alpha A^\beta A_\beta\Big] .
\eea
In order to put (\ref{SF}) into the functional form of (\ref{ChernSimons}), we note that, on the one hand, the trace over the generators $T^a$ of the Lie algebra is invariant under cyclic permutations. Therefore, by using this property, we are able to prove that
\bea
\label{id1}
\textrm{tr}\big(\{A^\alpha,A^\beta\}D_\beta A_\alpha\big)&=&\textrm{tr}\Big(\frac{3}{2}A^\alpha[A^\beta,D_\beta A_\alpha]-\frac{1}{2}\{A^\alpha,A^\beta\}D_\beta A_\alpha\Big) ;\\
\label{id2}
\frac{1}{3}\textrm{tr}\big(A^\alpha A^\beta A_\alpha A_\beta+A^\alpha A_\alpha A^\beta A_\beta\big)&=&\textrm{tr}\Big(\frac{1}{3}A^\alpha[A^\beta,\{A_\beta,A_\alpha\}]-\frac{1}{6}\{A^\alpha,A^\beta\}\{A_\alpha,A_\beta\}\Big).
\eea
On the other hand, in three dimensions, any object with three spinor indices satisfies the identity $O_{[\alpha\beta\gamma]}=0$, due to the fact that spinor indices can take only two values $\alpha=1,2$. Therefore, contracting $A_{[\alpha} A_\beta A_{\gamma]}=0$ with $C^{\beta\gamma}$, we are able to show that
\bea
\label{id3}
A^\beta A_\alpha A_\beta=A_\alpha A^\beta A_\beta+A^\beta A_\beta A_\alpha\Rightarrow\textrm{tr}(A^\alpha A^\beta A_\alpha A_\beta)=2\textrm{tr}(A^\alpha A_\alpha A^\beta A_\beta) \ .
\eea
By means of the identities (\ref{id1}-\ref{id3}), it is trivial to rewrite Eq. (\ref{SF}) as
\bea
\label{FinalChernSimons}
\Gamma_{odd}[A_\alpha]=-\frac{g^2}{8\pi|m|}\textrm{tr}\int d^5z\frac{m}{2g^2}\big(A^\alpha W_\alpha+\frac{i}{6}\{A^\alpha,A^\beta\}D_\beta A_\alpha+\frac{1}{12}\{A^\alpha,A^\beta\}\{A_\alpha, A_\beta\}\big) \ ,
\eea
where the Yang-Mills field strength superfield $W_\alpha$ is given by (\ref{strengthfield}). Of course, Eq. (\ref{FinalChernSimons}) is the non-Abelian Chern-Simons action \cite{Superbooks}, up to the constant factor $-\frac{g^2}{8\pi|m|}$. As a result, the gauge superfield $A_\alpha$ being non-dynamical on the classical level acquires a nontrivial dynamics due to quantum corrections. It is worth to point out that the methodology presented here can be easily adapted and applied to the noncommutative and Lorentz-breaking aether superspaces \cite{aether}.

\section{SUMMARY}

We succeeded to generate the non-Abelian supersymmetric Chern-Simons term. Within our calculations we, first, used the superfield formalism at all steps, second, did not impose any restrictions on structure of the gauge group, which makes our results to be universal, applicable to an arbitrary Lie algebra. As a consequence, we generalized the results of \cite{BDP,Dunne}, where the non-Abelian Chern-Simons term has been generated from the fermionic determinant, in a non-supersymmetric theory. The finiteness (and hence independence on any renormalization scheme) of our result is natural due to well-known one-loop finiteness of three-dimensional field theory models.

The most important consequence of our result is its natural interpretation within the context of the concept of the emergent dynamics.  Essentially, we have showed that the integration over scalar matter coupled to the external non-Abelian gauge superfield will yield the correct form of the non-Abelian supersymmetric Chern-Simons term. Therefore, we hope, that the approach we use in this paper can serve for more sophisticated supersymmetric field theory models, in particular, to generate the complete non-Abelian super-Yang-Mills action whose expression will involve Feynman supergraphs up to sixth order.

\vspace*{4mm}

{\bf Acknowledgments.} This work was partially supported by Conselho
Nacional de Desenvolvimento Cient\'{\i}fico e Tecnol\'{o}gico (CNPq) and Coordena\c c\~{a}o de Aperfei\c coamento de Pessoal de N\'{i}vel Superior (CAPES). The work by A. Yu. P. has been supported by the CNPq project No. 303438/2012-6.

\end{document}